\documentclass[10pt,twocolumn,letterpaper]{article}

\usepackage{cvpr}              % To produce the CAMERA-READY version
\usepackage{graphicx}
\usepackage{amsmath}
\usepackage{amssymb}
\usepackage{booktabs}
\usepackage{multirow}
\usepackage[pagebackref,breaklinks,colorlinks]{hyperref}

\usepackage[capitalize]{cleveref}
\crefname{section}{Sec.}{Secs.}
\Crefname{section}{Section}{Sections}
\Crefname{table}{Table}{Tables}
\crefname{table}{Tab.}{Tabs.}

\def\cvprPaperID{5882} % *** Enter the CVPR Paper ID here
\def\confName{CVPR}
\def\confYear{2022}

\begin{document}
	
	%\title{DuDoTransformer: Dual-Domain \\Transformer for Sparse-View CT Reconstruction}
	
	\title{
		DuDoTrans: Dual-Domain Transformer Provides More \\Attention for Sinogram Restoration in Sparse-View CT Reconstruction}
	
	\author{Ce Wang$^{1,2}$\quad
	Kun Shang$^{3}$\quad
	Haimiao Zhang$^{4}$\quad
	Qian Li$^{1,2}$\quad
	Yuan Hui$^{1,2}$\quad
	S. Kevin Zhou$^{5,*}$\quad\\
	$^{1}$Institute of Computing Technology, CAS, Beijing, China\\
	$^{2}$Suzhou Institute of Intelligent Computing Technology, CAS, Suzhou, China\\
	$^{3}$Shenzhen Institutes of Advanced Technology, CAS, Shenzhen, China \\
%	Research Center for Medical AI, 
	$^{4}$Beijing Information Science and Technology University, Beijing, China\\
%	$^{5}$Medical Imaging, Robotics, and Analytic Computing Laboratory and Engineering (MIRACLE) Center, \\ 
%	School of Biomedical Engineering and Suzhou Institute for Advanced Research, \\
	$^{5}$University of Science and Technology of China, Suzhou, China \\
	{\tt\small s.kevin.zhou@gmail.com}
	%% omit the following lines up until the closing ``}''.
	%% Additional authors and addresses can be added with ``\and'',
	%% just like the second author.
	%% To save space, use either the email address or home page, not both
%	\and
%	Kun Shang\\
%	Research Center for Medical AI, Shenzhen Institutes
%	of Advanced Technology, CAS, Shenzhen, China \\
%	{\tt\small secondauthor@i2.org}
	}
	\maketitle
	
	%%%%%%%%% ABSTRACT
	\begin{abstract}
		While Computed Tomography (CT) reconstruction from X-ray sinograms is necessary for clinical diagnosis, iodine radiation in the imaging process induces irreversible injury, thereby driving researchers to study sparse-view CT reconstruction, that is, recovering a high-quality CT image from a sparse set of sinogram views. Iterative models are proposed to alleviate the appeared artifacts in sparse-view CT images, but the computation cost is too expensive. Then deep-learning-based methods have gained prevalence due to the excellent performances and lower computation. However, these methods ignore the mismatch between the CNN's \textbf{local} feature extraction capability and the sinogram's \textbf{global} characteristics. To overcome the problem, we propose \textbf{Du}al-\textbf{Do}main \textbf{Trans}former (\textbf{DuDoTrans}) to simultaneously restore informative sinograms via the long-range dependency modeling capability of Transformer and reconstruct CT image with both the enhanced and raw sinograms. With such a novel design, reconstruction performance on the NIH-AAPM dataset and COVID-19 dataset experimentally confirms the effectiveness and generalizability of DuDoTrans with fewer involved parameters. Extensive experiments also demonstrate its robustness with different noise-level scenarios for sparse-view CT reconstruction. The code and models are publicly available at \href{https://github.com/DuDoTrans/CODE}{https://github.com/DuDoTrans/CODE}.
		
	\end{abstract}
	
	%%%%%%%%% BODY TEXT
	\section{Introduction and Motivation}\label{sec:intro}
	
	Computed Tomography (CT) is a widely used clinically diagnostic imaging procedure aiming to reconstruct a clean CT image $\mathbf{X}$ from observed sinograms $\mathbf{Y}$, but its accompanying radiation heavily limits its practical usage. To decrease the induced radiation dose and reduce the scanning time, Sparse-View (SV) CT is commonly applied. However, the deficiency of projection views brings severe artifacts in the reconstructed images, especially when common reconstruction methods such as analytical Filtered Backprojection (FBP) and algebraic reconstruction technique (ART)~\cite{natterer2001mathematics} are used, which poses a significant challenge to image reconstruction.
	
	\begin{figure}[t]
		\begin{center}
			\includegraphics[height=6.5cm]{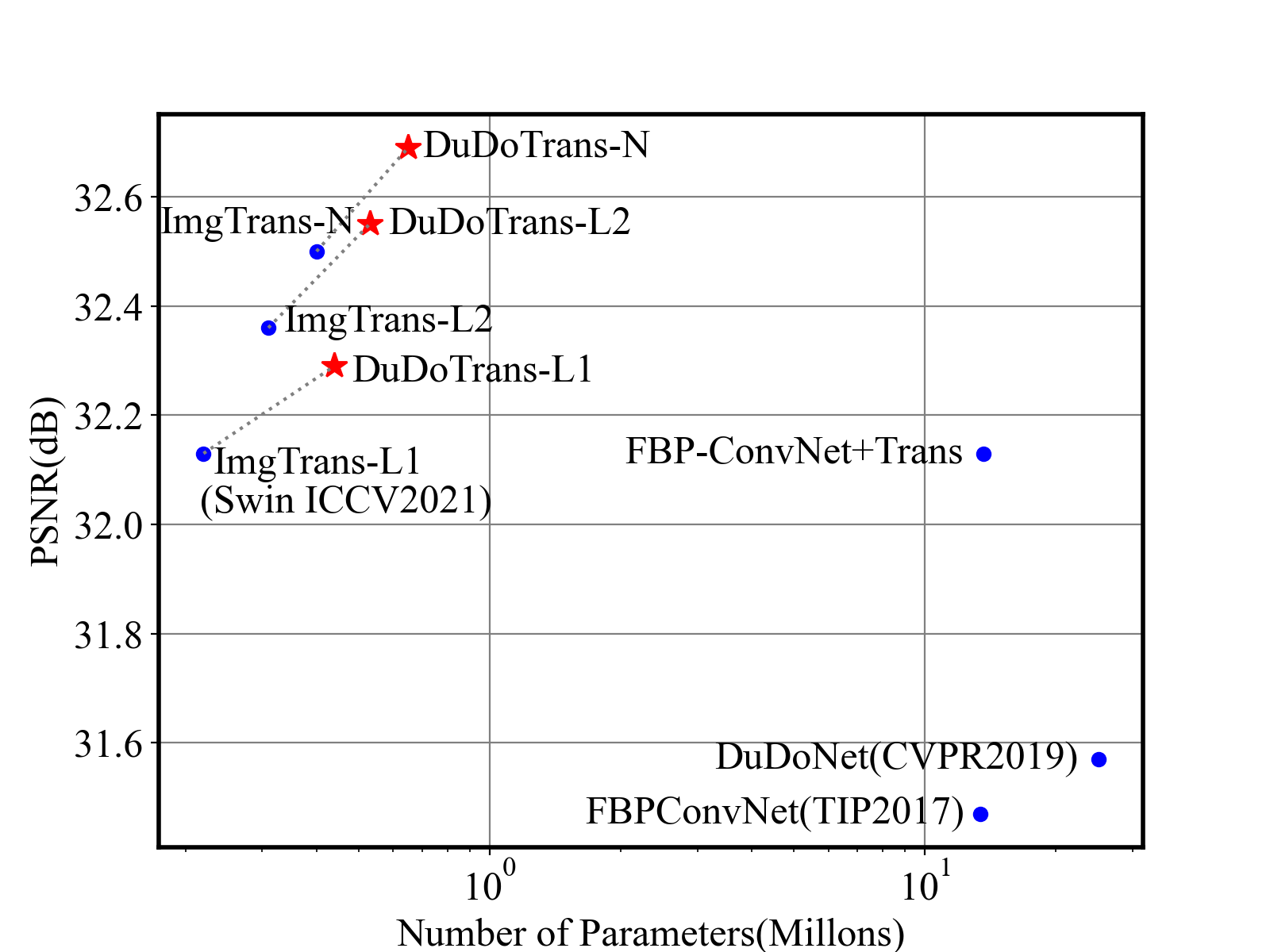}
			\caption{ Parameters versus performances of DuDoTrans and deep-learning-based CT reconstruction methods. L1 and L2 are two light versions while N represents the normal version. Firstly, we find Transformer-based reconstruction methods have consistently achieved \textbf{better performance with fewer parameters}. Further, our DuDoTrans derives better results than at a less computation cost.}			\label{performance_params}
			\vspace{-25pt}
		\end{center}
	\end{figure}
	
	To tackle the artifacts, some iterative methods are proposed to impose the well-designed prior knowledge (ideal image properties) via additional regularization terms $R(\mathbf{X})$, such as Total Variation (TV) based methods~\cite{sidky2008image, mahmood2018adaptive}, Nonlocal-based methods~\cite{zeng2015spectral}, and sparsity-based methods~\cite{bao2019convolutional, kim2014sparse}. Although these models have achieved better qualitative and quantitative performances, they suffer from over-smoothness. Besides, the iterative optimization procedure is often computationally expensive and requires careful case-by-case hyperparameter tuning, which is practically less applicable.
	
	With the success of CNNs in various vision tasks ~\cite{he2016deep,xiu2020supercharging,huang2017densely,zhou2021review,zhou2021deep}, CNN-based models are carefully designed and exhibit potential to a fast and efficient CT image reconstruction~\cite{chen2017low,jin2017deep,yang2018low,adler2018learned,cheng2020learned,gupta2018cnn}. These methods render the potential to learn a better mapping between low-quality images, such as reconstructed results of FBP, and ground-truth images. Recently, Vision Transformer~\cite{carion2020end, chen2021pre, dosovitskiy2020image, li2021localvit} has gained attention with its long-range dependency modeling capability, and numerous models have been proposed in medical image analysis~\cite{chen2021pre,zhou2021nnformer,cao2021swin,zhang2021learning,yu2021mil}. For example, TransCT~\cite{zhang2021transct} is proposed as an efficient method for low-dose CT reconstruction, while it suffers from memory limitation with involved patch-based operations. Besides, these deep learning-based methods ignore the informative sinograms, which makes their reconstruction inconsistent with the observed sinograms.
	
	To alleviate the problem, a series of dual-domain (DuDo) reconstruction models~\cite{lin2019dudonet,wang2021improving,zhou2020dudornet,zhou2021dudodr} are proposed to simultaneously enhance raw sinograms and reconstruct CT images with both enhanced and raw sinograms, experimentally showing that enhanced sinograms contribute to the latter reconstruction. Although these DuDo methods have shown satisfactory performances, \textbf{they neglect the global nature of the sinogram's sampling process, which is inherently hard to be captured via CNNs, as CNNs are known for extracting local spatial features. This motivates us to go a step further and design a more applicable architecture for sinogram restoration.}
	
	Inspired by the long-range dependency modeling capability \& shifted window self-attention mechanism of Swin Transformer~\cite{liu2021swin}, we specifically design the Sinogram Restoration Transformer (SRT) by considering the time-dependent characteristics of sinograms, which restore informative sinograms and overcome the mismatch between the global characteristics of sinograms and local feature modeling of CNNs.	
	Based on the SRT module, we finally propose \textbf{Du}al-\textbf{Do}main \textbf{Trans}former (\textbf{DuDoTrans}) to reconstruct CT image. Compared with previous image reconstruction methods, we summarize several benefits of DuDoTrans as follows:
	\vspace{-6pt}
	
	\begin{itemize}
		\item Considering the global sampling process of sinograms, we introduce SRT module, which has the advantages of both Swin-Transformer and CNNs. It has the desired long-range dependency modeling ability, which helps better restore the sinograms and has been experimentally verified in CNN-based, Transformer-based, and deep-unrolling-based reconstruction framework.
		
		\vspace{-6pt}
		\item With the powerful SRT module for sinogram restoration, we further propose Residual Image Reconstruction Module (RIRM) for image-domain reconstruction. To compensate for the drift error between the dual-domain optimization directions, we finally utilize the proposed differentiable DuDo Consistency Layer to keep the restored sinograms consistent with reconstructed CT images, which induces the final DuDoTrans. Hence, DuDoTrans not only has the desired long-range dependency and local modeling ability, but also has the benefit of dual-domain reconstruction.
		
		\vspace{-6pt}		
		\item Reconstruction performance on the NIH-AAPM dataset and COVID-19 dataset experimentally confirms the effectiveness, robustness, and generalizability of the proposed method. Besides, by adaptively employing Swin-Transformer and CNNs, our DuDoTrans has achieved better performance with fewer parameters as shown in Figure~\ref{performance_params} and similar FLOPs (shown in later experiments), which makes the model practical in various applications.
		%but also impose sinogram prior knowledge from projection domain to the image domain
	\end{itemize}
	%simultaneously restore clean sinograms with SRT and reconstruct CT image with both raw and enhanced sinograms
	
	\section{Backgrounds and Related Works}
	\subsection{Tomographic Image Reconstruction}
	Human body tissues, such as bones and organs, have different X-ray attenuation coefficients $\mu$. When considering a 2D CT image, the distribution of the attenuation coefficients $\mathbf{X}=\mu(a,b)$, where $(a,b)$ indicate positions,  represents the underlying anatomical structure. The principle of CT imaging is based on the fundamental Fourier Slice Theorem, which guarantees that the 2D image function $\mathbf{X}$ can be reconstructed from the obtained dense projections (called sinograms). When imaging, projections of the anatomical structure $\mathbf{X}$ are indeed inferred by the emitted and received X-ray intensities according to the Lambert-Beer Law. Further, when under a polychromatic X-ray source with an energy distribution $\eta(E)$, the CT imaging process is given as:
	\begin{equation}
	\mathbf{Y}=-\log \int \eta(E) \exp \{ -\mathcal{P\mathbf{X}} \},
	\end{equation}
	where $\mathcal{P}$ represents the sinogram generation process, i.e., Radon transformation (commonly defined with a fan-beam imaging geometry). With the above forward process, CT imaging aims to reconstruct $\mathbf{X}$ from the obtained projections $\mathbf{Y}=\mathcal{P}\mathbf{X}$ (abbreviation for simplicity) with the estimated/learned function ${\mathcal{P}}^{\dagger}$. In practical SVCT, the projection data $\mathbf{Y}$ is incomplete, where the total $\alpha_{max}$ projection views are sampled uniformly in a circle around the patient. This reduced sinogram information heavily limits the performance of previous methods and results in artifacts. In order to alleviate the phenomena, many works have been recently proposed, which can be categorized into the following two groups: Iterative-based reconstruction methods~\cite{sidky2008image, mahmood2018adaptive,zeng2015spectral,bao2019convolutional, kim2014sparse} and Deep-learning-based reconstruction methods\cite{he2016deep,xiu2020supercharging,huang2017densely,zhou2021review,zhou2021deep}. Different from these prevalent works, \textbf{our method DuDoTrans is based on deep-learning, but is the only dual-domain method based on Transformer.}
	
	\begin{figure*}[h]
		\begin{center}
			\includegraphics[height=6.2cm]{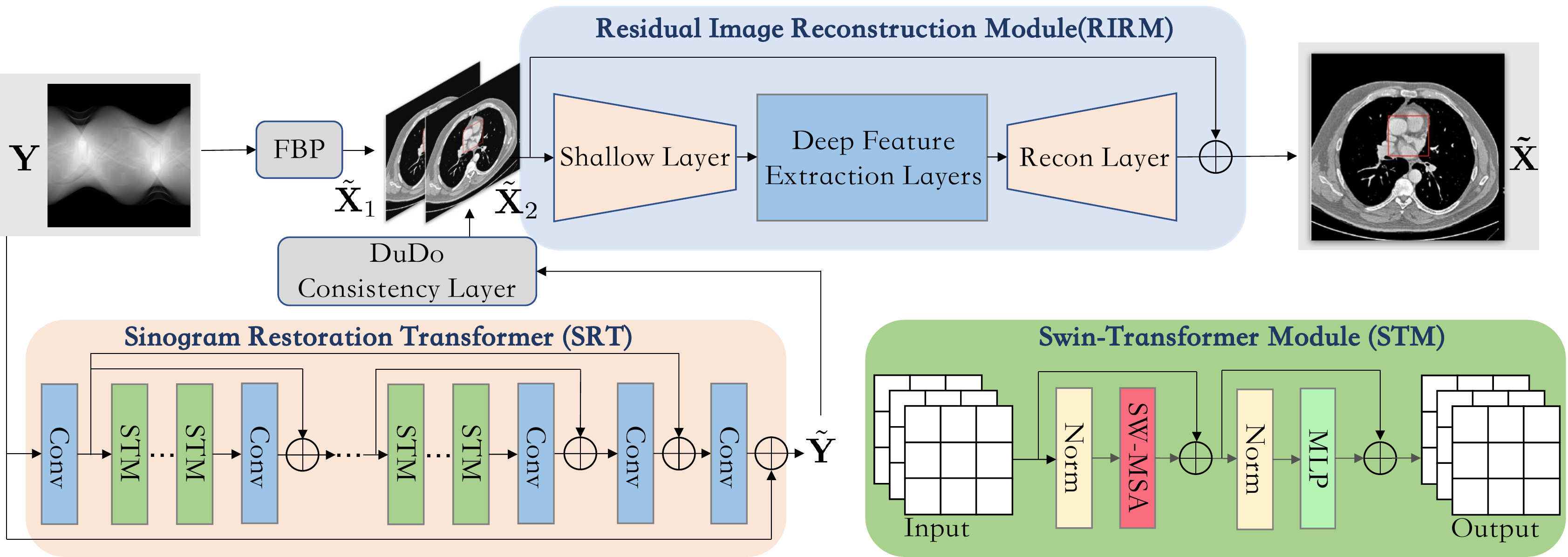}
			\caption{The framework of proposed DuDoTrans for SV CT image reconstruction. When under-sampled sinograms are given, our DuDoTrans first restores clean sinograms with SRT, followed by RIRM to reconstruct the CT image with both restored and raw sinograms.}
			\label{framework}
			\vspace{-20pt}
		\end{center}
	\end{figure*}

	%In this case, the problem becomes very ill-posed, and regularization-based methods~\cite{sidky2008image, mahmood2018adaptive,zeng2015spectral,bao2019convolutional, kim2014sparse} have been adopted and achieved satisfactory performance. They introduce expert designed regularization terms $R(\mathbf{X})$ to impose ideal image properties. Nevertheless, they suffer from over-smoothness. Besides, the iterative optimization procedure imposes an expensive computation cost, which is practically less applicable.
	%With the success of deep learning in vision tasks~\cite{he2016deep,xiu2020supercharging,huang2017densely,zhou2021review,zhou2021deep} and the release of medical image datasets, deep-learning-based reconstruction methods~\cite{chen2017low,jin2017deep,yang2018low, zhang2018sparse, han2018framing,lee2018deep} renders the potential to learn a better mapping between low-quality images, such as reconstructed results of FBP, and ground-truth images. But the ignoring of sinogram information in these methods makes their reconstruction $\mathbf{\hat{X}}$ inconsistent with the observed $\mathbf{Y}$. Therefore, a series of dual-domain reconstruction works~\cite{lin2019dudonet,wang2021improving,zhou2020dudornet,zhou2021dudodr,lyu2020encoding} have been successfully proposed to solve the problem, experimentally showing that the enhanced sinograms contributes to the reconstruction. 
	
	\subsection{Transformer in Medical Imaging}
	Based on the powerful attention mechanism~\cite{vaswani2017attention,cao2019gcnet,fu2019dual,yin2020disentangled,ramachandran2019stand,zhao2020exploring} and patch-based operations, Transformer is applied to many vision tasks~\cite{dosovitskiy2020image, touvron2021training, han2021transformer, liu2021swin}. Especially, Swin Transformer~\cite{liu2021swin} incorporates such an advantage with the local feature extraction ability of CNNs. With such an intuitive manner, Swin-Transformer-based models~\cite{liang2021swinir} have relieved the limitation of memory in previous Vision Transformer-based models. Based on these successes and for better modeling the global features of medical images, Transformer has been applied to medical image segmentation~\cite{chen2021pre,zhou2021nnformer,cao2021swin,lin2021ds}, registration~\cite{zhang2021learning}, classification~\cite{yu2021mil,wang2021transpath}, and achieved surprising improvements. Nevertheless, few works explore Transformer structures in SVCT reconstruction. Although TransCT~\cite{zhang2021transct} attempts to suppress the noise artifacts in low-dose CT with Transformer, \textbf{they neglect the consideration of global sinogram characteristics in their design, which is taken into account in DuDoTrans.}
	
	\section{Method}
	As shown in Fig.~\ref{framework}, we build DuDoTrans with three modules: (a) Sinogram Restoration Transformer (SRT), (b) DuDo Consistency Layer, and (c) Residual Image Reconstruction Module (RIRM). Assume that a sparse-view sinogram $\mathbf{Y}\in \mathcal{R}^{{H_s}\times{W_s}}$ is given, we first use FBP~\cite{natterer2001mathematics} to reconstruct a low-quality CT image $\tilde{\mathbf{{X}}}_{1}$. Simultaneously, the SRT module is introduced to output an enhanced sinogram $\tilde{\mathbf{Y}}$, followed by the DuDo Consistency Layer to yield another estimation $\tilde{\mathbf{{X}}}_{2}$. At last, these low-quality images $\tilde{\mathbf{{X}}}_{1}$ and $\tilde{\mathbf{{X}}}_{2}$ are concatenated and fed into RIRM to predict the CT image $\mathbf{\tilde{X}}$, which will be supervised with the corresponding clean CT image $\mathbf{X}_{gt}\in  \mathcal{R}^{{H_I}\times{W_I}}$. We next introduce the above-involved modules in detail.
	
	\subsection{Sinogram Restoration Transformer}
	\label{srt}
	Sinogram restoration is extremely challenging since the intrinsic information not only contains spatial structures of human bodies, but follows the global sampling process. Specifically, each line $\{\mathbf{{Y}_{i}}\}_{i=1}^{{H}_{s}}$ of a sinogram $\mathbf{Y}$ are sequentially sampled with overlapping information of surrounding sinograms. In other words, 1-D components of sinograms heavily correlate with each other. The global characteristic makes it difficult to be captured with traditional CNNs, which are powerful in local feature extraction. For this reason, we equip this module with the Swin-Transformer structure, which enables it with long-range dependency modeling ability. As shown in Fig.~\ref{framework}, SRT consists of $m$ successive residual blocks, and each block contains $n$ normal Swin-Transformer Module (STM) and a spatial convolutional layer, which have the capacity of both global and local feature extraction. Given the degraded sinograms, we first use a convolutional layer to extract the spatial structure $\mathbf{F}_{conv} $. Considering it as $\mathbf{F}_{{STM}_{0}}$, then  $n$ STM components of each residual block output $\{\mathbf{F}_{{STM}_{i}}\}_{i=1}^{m}$ with the following formulation:
	\begin{equation}
	\mathbf{F}_{{STM}_{i}} = \mathnormal{M}_{conv}( \prod_{j=1}^{n} \mathnormal{M}_{swin}^{j} ( \mathbf{F}_{{STM}_{i-1}} ) ) + \mathbf{F}_{{STM}_{i-1}},
	\end{equation}
	where $\mathnormal{M}_{conv}$ denotes a convolutional layer, $\{\mathnormal{M}_{swin}^{j}\}_{j=1}^{n}$ denotes $n$ Swin-Transformer layers, and $\prod$ represents the successive operation of Swin-Transformer Layers. Finally, the enhanced sinograms are estimated with:
	\begin{equation}
	\tilde{\mathbf{Y}} = \mathbf{Y} + \mathnormal{M}_{conv}(\mathnormal{M}_{conv}(\mathbf{F}_{{STM}_{m}}) + \mathbf{F}_{{STM}_{0}}).
	\end{equation}
	As a restoration block, ${\mathcal{L}}_{SRT}$ is used to supervise the output of the SRT:
	\begin{equation}
	\mathcal{L}_{SRT} = \| \mathbf{\tilde{Y}} - \mathbf{Y}_{gt} \|_{2},
	\end{equation}
	where $\mathbf{Y}_{gt}$ is the ground truth sinogram, and it should be given when training.
	
	\subsection{DuDo Consistency Layer}
	Although input sinograms have been enhanced via the SRT module, directly learning from the concatenation of $\tilde{\mathbf{{X}}}_{1}$ and $\tilde{\mathbf{{X}}}_{2}$ leaves a drift between the optimization directions of SRT and RIRM. To compensate for the drift, we make use of a differentiable DuDo Consistency Layer ${\mathnormal{M}}_{DC}$ to back-propagate the gradients of RIRM. In this way, the optimization direction imposes the preferred sinogram characteristics to $\mathbf{\tilde{Y}}$, and vice versa. To be specific, given the input fan-beam sinogram $\mathbf{\tilde{Y}}$, the DuDo Consistency Layer first converts it into parallel-beam geometry, followed with Filtered Backprojection:
	\begin{equation}
	\tilde{\mathbf{{X}}}_{2} = {\mathnormal{M}}_{DC}( \mathbf{\tilde{Y}}).
	\end{equation}
	To additionally keep the restored sinograms consistent with the ground-truth CT image $\mathbf{X}_{gt}$, ${\mathcal{L}}_{DC}$ is proposed as follows:
	\begin{equation}
	\mathcal{L}_{DC} = \| \tilde{\mathbf{{X}}}_{2} - \mathbf{X}_{gt} \|_{2}.
	\end{equation}
	
	\subsection{Residual Image Reconstruction Module}
	As a long-standing clinical problem, the final goal of CT image reconstruction is to recover a high-quality CT image for diagnosis. With the initially estimated low-quality images that help rectify the geometric deviation between the sinogram and image domains, we next employ Shallow Layer ${\mathnormal{M}}_{sl}$ to obtain shallow features of input low-quality image $\mathbf{\tilde{X}}$:
	\begin{equation}
	\mathbf{F}_{sl} = {\mathnormal{M}}_{sl} ([\tilde{\mathbf{{X}}}_{1}, \tilde{\mathbf{{X}}}_{2}]). %\text{where} [] \text{means concat}.
	\end{equation}
	Then a series of Deep Feature Extraction Layers ${\{\mathnormal{M}}_{df}^{i}\}_{i=1}^{n}$ are introduced to extract deep features:
	\begin{equation}
	\mathbf{F}_{df}^{i} = \mathnormal{M}_{df}^{i} (\mathbf{F}_{df}^{i-1}), ~i=1,2,\ldots,n,
	\end{equation}
	where $\mathbf{F}_{df}^{0}=\mathbf{F}_{sl}$. Finally, we utilize a Recon Layer ${\mathnormal{M}}_{re}$ to predict the clean CT image with residual learning:
	\begin{equation}
	\mathbf{\tilde{X}} = {\mathnormal{M}}_{re} (\mathbf{F}_{df}^{n}) + \tilde{\mathbf{{X}}}_{1}.
	\end{equation}
	To supervise our network optimization, the below ${\mathcal{L}}_{RIRM}$ loss is used for this module:
	\begin{equation}
	\mathcal{L}_{RIRM} = \| \mathbf{\tilde{X}} - \mathbf{X}_{gt} \|_{2}.
	\end{equation}
	The full objective of our model is:
	\begin{equation}
	\mathcal{L} = \mathcal{L}_{SRT} + {\lambda}_{1}\mathcal{L}_{DC} + {\lambda}_{2}\mathcal{L}_{RIRM},
	\end{equation}
	where ${\lambda}_{1}$ and ${\lambda}_{2}$ are blending coefficients, which are both empirically set as 1 in experiments.
	
	Note that the intermediate convolutional layers are used to communicate between image space $ \mathcal{R}^{{H_I}\times{W_I}}$ and patch-based feature space $ \mathcal{R}^{\frac{{H_I}}{w}\times\frac{{W_I}}{w}\times {w}^{2} }$. Further, by dynamically tuning the depth $m$ and width $n$, SRT modules are flexible in practice depending on the balance between memory and performance. We will explore this balancing issue in later experiments.
	
	\section{Experimental Results}
	\subsection{Experimental Setup}
	
	\noindent\textbf{Datasets.} We first train and test our model with the ``2016 NIH-AAPM-Mayo Clinic Low Dose CT Grand Challenge"~\cite{mccollough2016tu} dataset. Specifically, we choose a total of 1746 slices (resolution 512$\times$512) from five patients to train our models, and use 314 slices of another patient for testing. We employ a scanning geometry of Fan-Beam X-Ray source with 800 detector elements. There are four SV scenarios in our experiment, corresponding to ${\alpha}_{max}$ = [24, 72, 96, 144] views. Note that these views are uniformly distributed around the patient. The original dose data are collected from the chest to the abdomen under a protocol of 120 kVp and 235 effective mAs (500mA/0.47s). To simulate the photon noise in numerical experiments, we add to sinograms mixed noise that is by default composed of 5\% Gaussian noise and Poisson noise with an intensity of 5${e}^{\text{6}}$.
	
	\noindent\textbf{Implementation details and training settings.} Our models are implemented using the PyTorch framework. We use the Adam optimizer~\cite{kingma2014adam} with $({\beta}_{1}, {\beta}_{2})$ = (0.9, 0.999) to train these models. The learning rate starts from 0.0001. Models are all trained on a Nvidia 3090 GPU card for 100 epochs with a batch size of 1.
	
	\noindent\textbf{Evaluation metrics.} Reconstructed CT images are quantitatively measured by the multi-scale Structural Similarity Index Metric (SSIM) (with level = 5, Gaussian kernel size = 11, and standard deviation = 1.5) ~\cite{wang2004image,wang2003multiscale} and Peak Signal-to-Noise Ratio (PSNR).	
	\begin{table}[t!]
		\begin{center}
			\caption{The effect of each module (CNNs v.s. Transformer) on the reconstruction performance. Obviously, our DuDoTrans design achieves the best performance.}
			\label{effect-srt}
			\begin{tabular}{  l |rrr } 
				\hline 
				Method                  &PSNR & SSIM & RMSE \\
				\hline
				
				FBPConvNet         & 31.47 & 0.8878 & 0.0268  \\
				DuDoNet      &31.57  &0.8920 &0.0266\\
				FBPConvNet+SRT&   32.13   & 0.8989 & 0.0248 \\
				ImgTrans  &  32.50&  0.9010 & 0.0238\\
				\textbf{DuDoTrans}    &\textbf{32.68} &\textbf{0.9047}  &\textbf{0.0233} \\
				\hline
			\end{tabular}
		\end{center}
		\vspace{-20pt}
	\end{table}
	
	\begin{figure*}[!t]
		\begin{minipage}[t]{0.32\textwidth}
			\centering
			\includegraphics[width=0.95\textwidth]{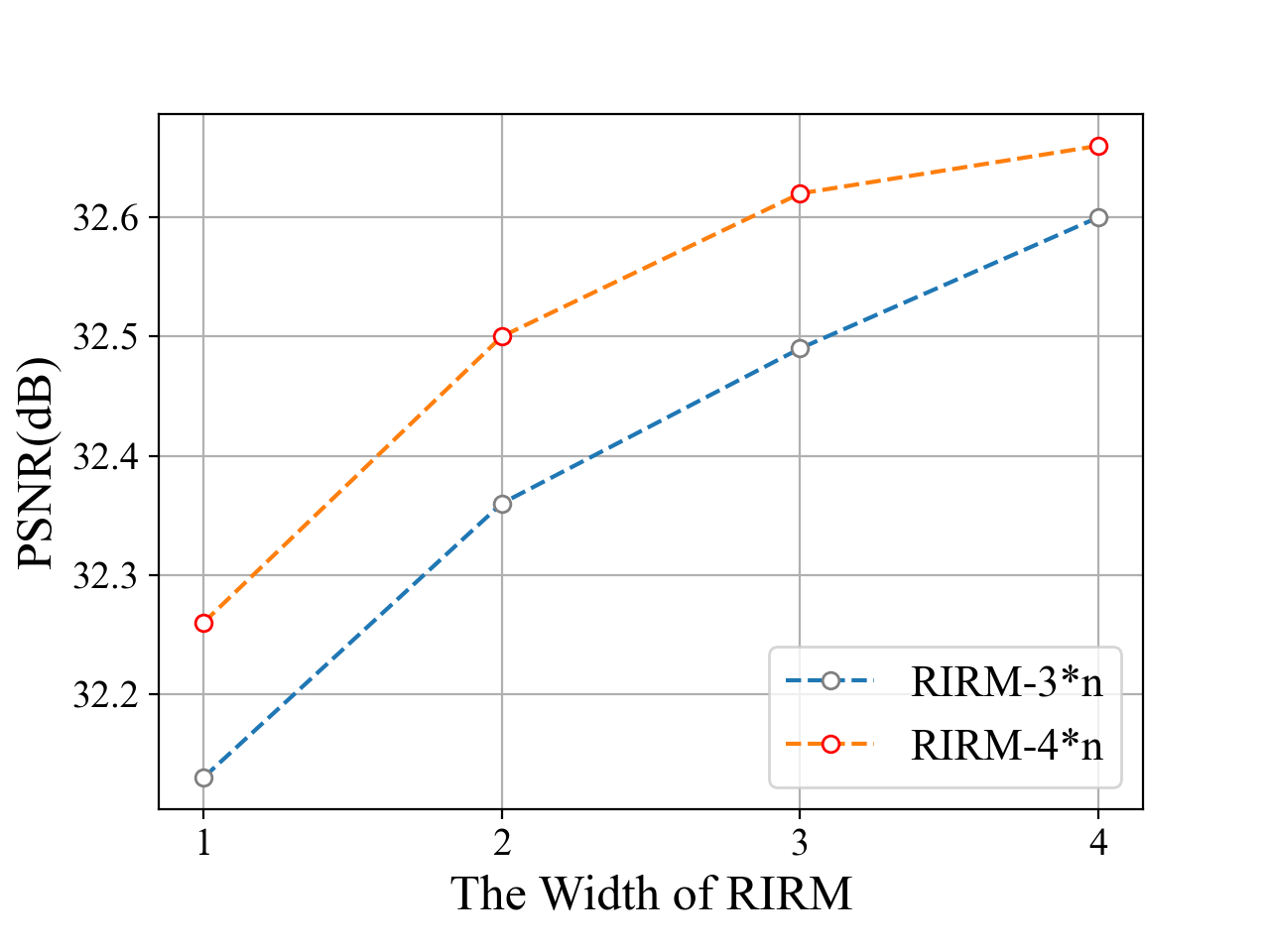}
			(a)
			\includegraphics[width=0.95\textwidth]{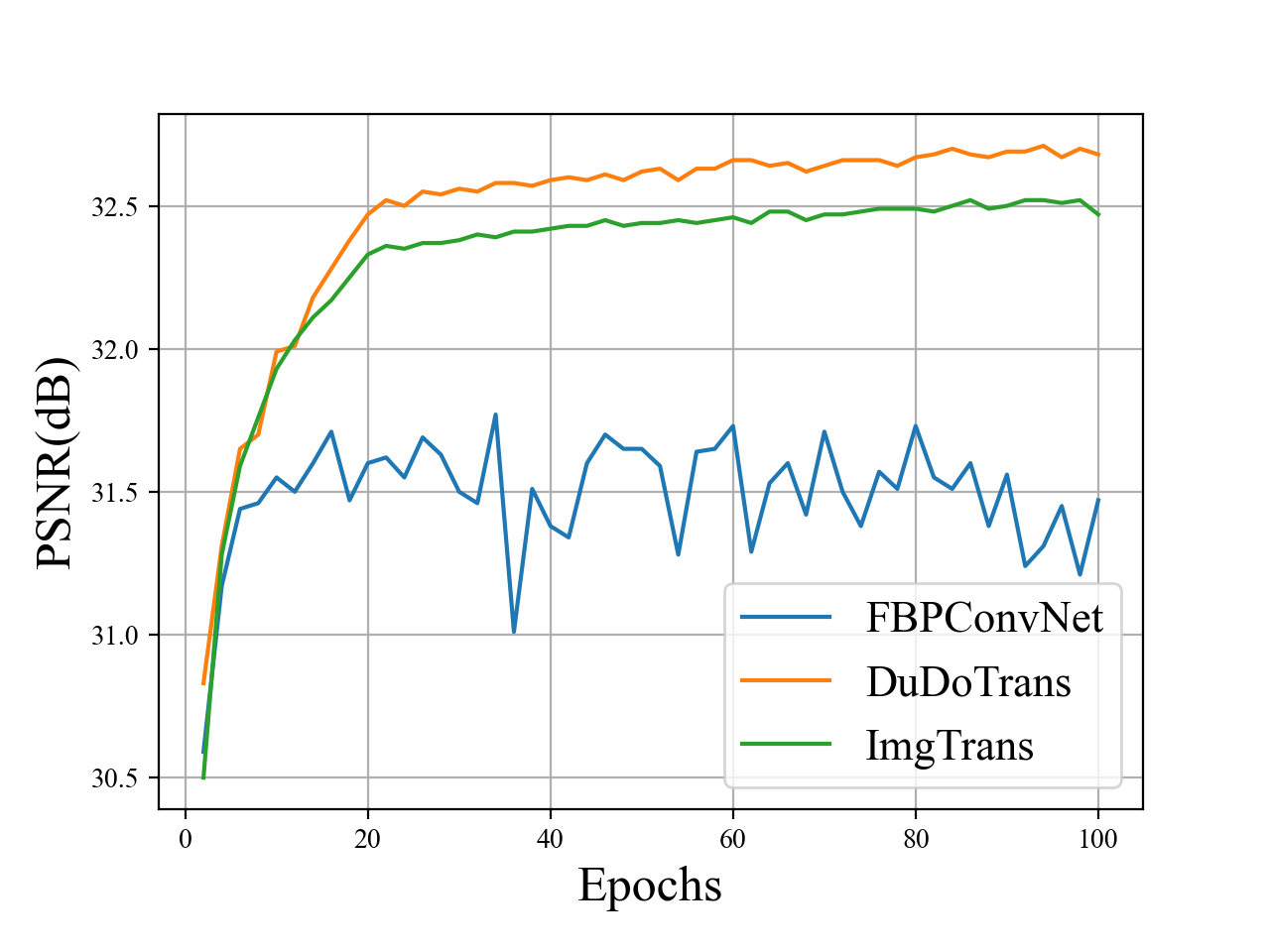}
			(d)
		\end{minipage}
		\begin{minipage}[t]{0.32\textwidth}
			\centering
			\includegraphics[width=0.95\textwidth]{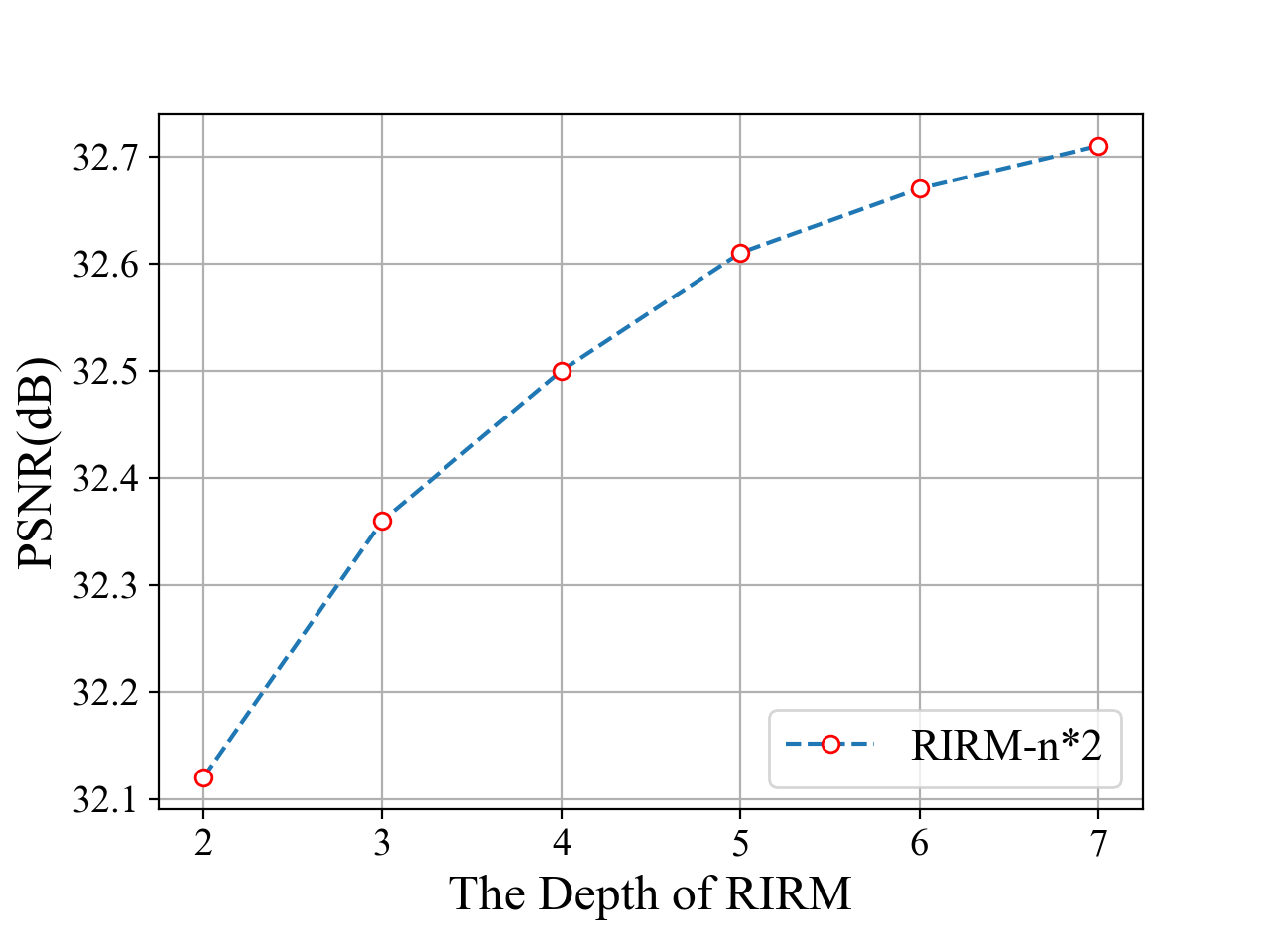}
			(b)
			\includegraphics[width=0.95\textwidth]{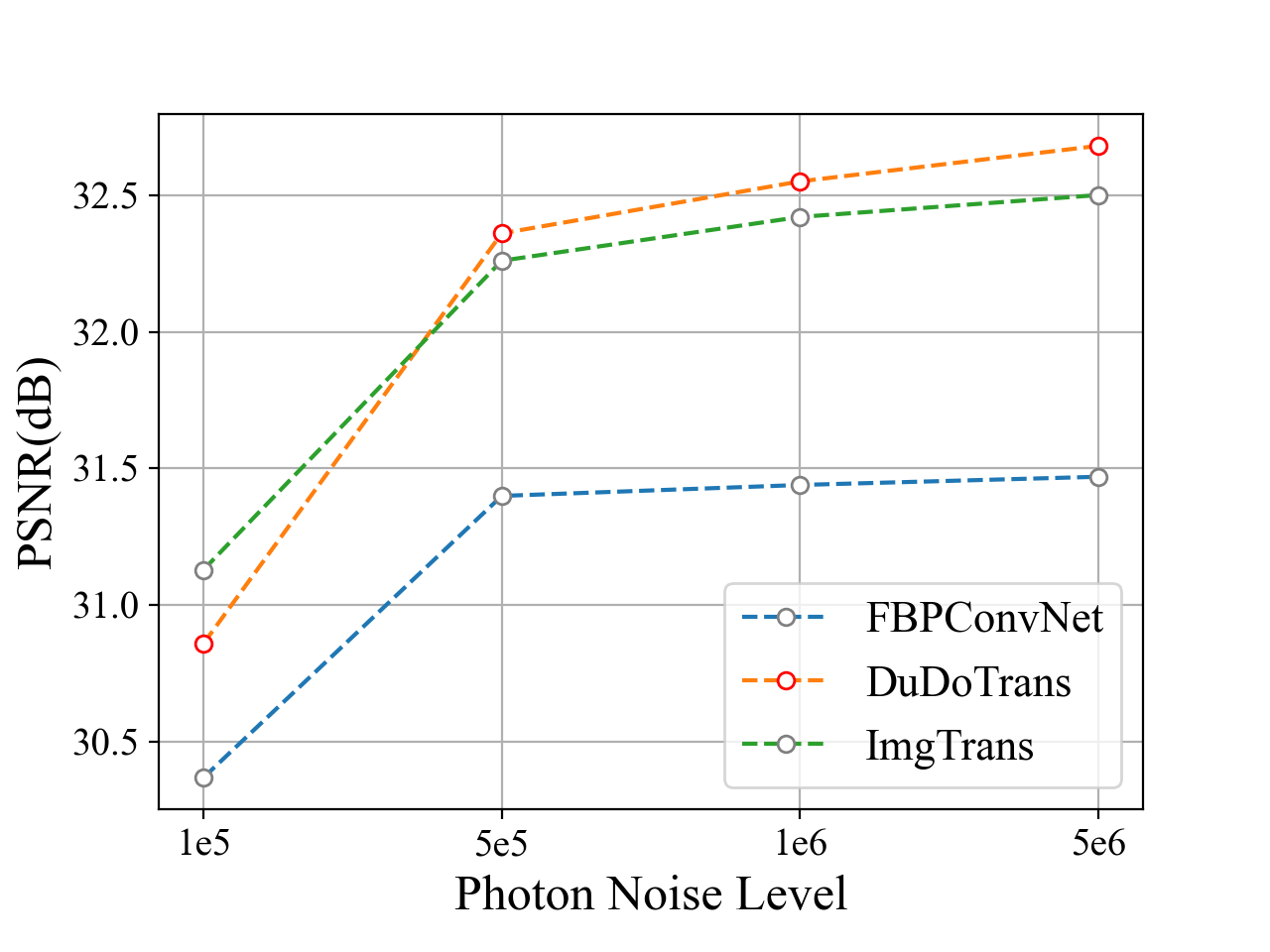}
			(e)
		\end{minipage}
		\begin{minipage}[t]{0.32\textwidth}
			\centering
			\includegraphics[width=0.95\textwidth]{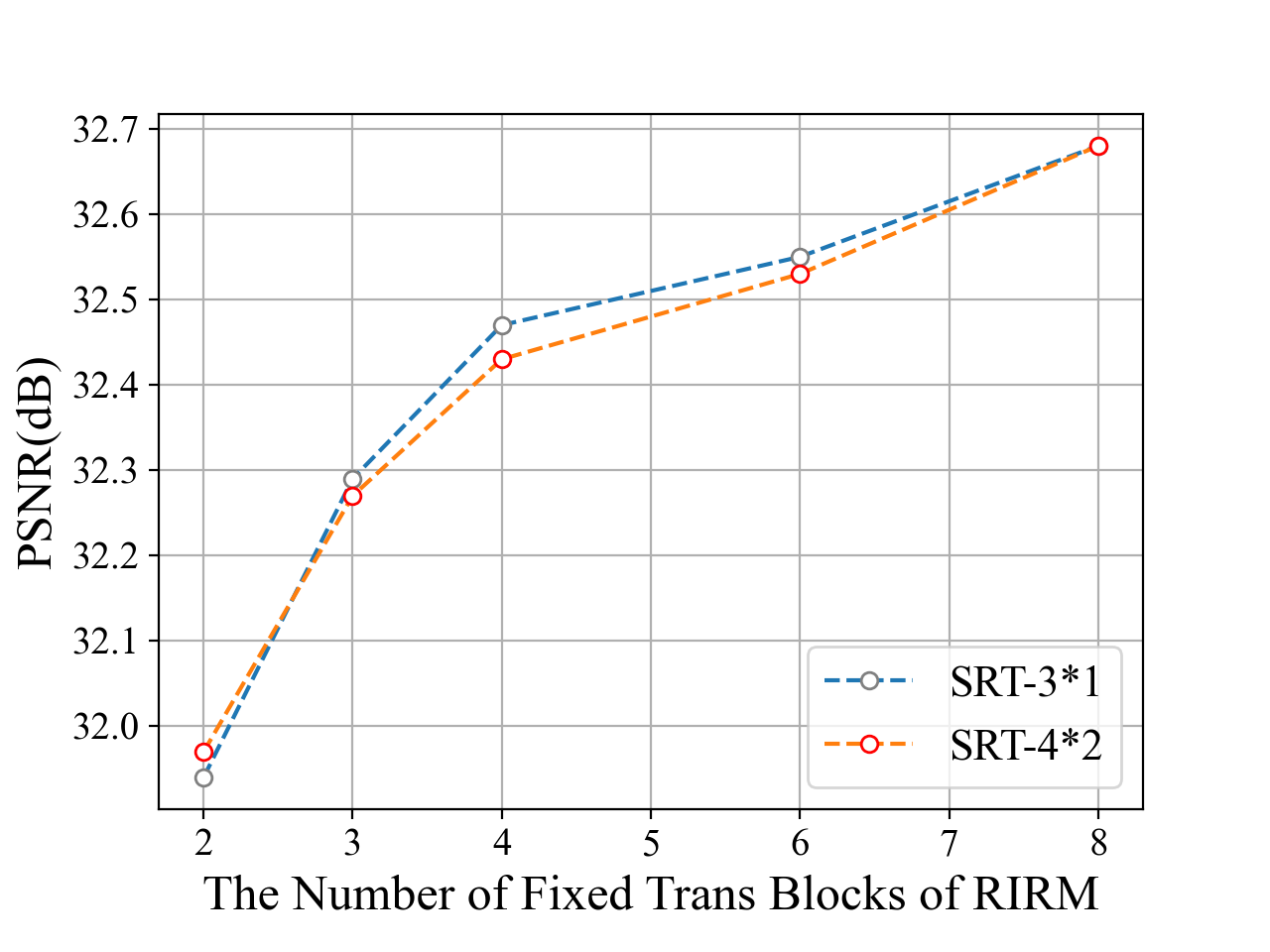}
			(c)
			\includegraphics[width=0.95\textwidth]{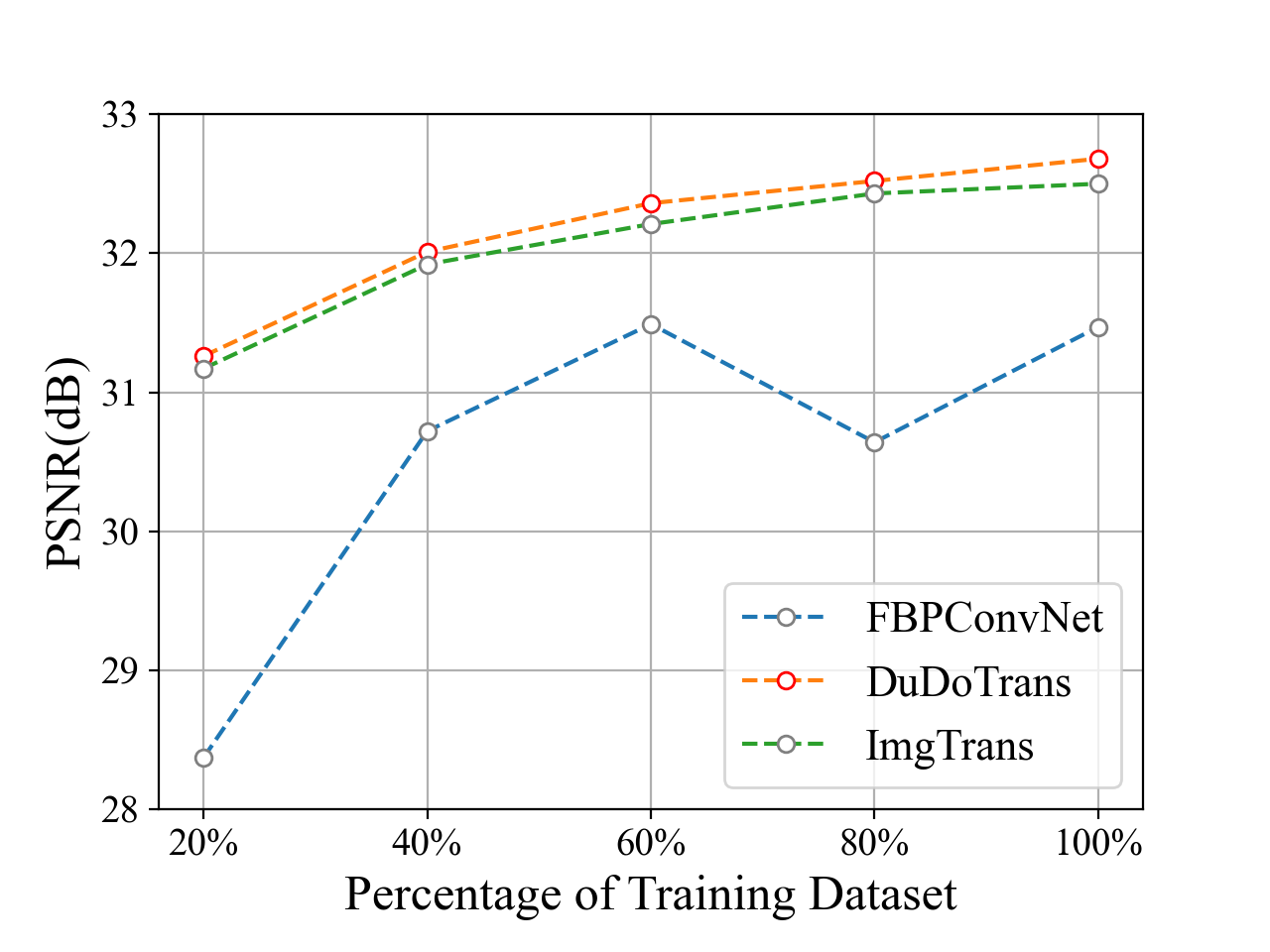}
			(f)
		\end{minipage}
		\caption{The first row compares the effect of RIRM depth, RIRM width, and SRT size on reconstruction. The second row inspects the convergence, robustness on noise, and the effect of training dataset scale on DuDoTrans.}
		\label{ablation}
		\vspace{-10pt}
	\end{figure*}
	
	\subsection{Ablation Study and Analysis}
	We next prove the effectiveness of our proposed SRT module and exhaust the best structure for DuDoTrans. Firstly, we conduct the experiments with the five models: (a) FBPConvNet~\cite{jin2017deep}, (b) DuDoNet~\cite{lin2019dudonet}, (c) FBPConvNet+SRT, which combines (a) with our proposed SRT, (d) ImgTrans, which replaces the image-domain model in (a) with Swin-Transformer~\cite{liu2021swin}, and (e) our DuDoTrans. The experimental settings are by default with ${\alpha}_{max}$ = 96, and the results are shown in Table~\ref{effect-srt}. \\	
	\noindent\textbf{The Effectiveness of SRT.} Comparing models (a) and (c) in Table~\ref{effect-srt}, the performance is improved 0.66 dB, which confirms that the SRT module output $\tilde{\mathbf{Y}}$ indeed provides useful information for the image-domain reconstruction. \\
	\noindent\textbf{The exploration of RIRM.} Inspired by the success of Swin-Transformer in low-level vision tasks, we simply replace the post-precessing module of FBPConvNet with Swin-Transformer, named ImgTrans. Comparing it with the baseline model (a), the achieved 1 dB improvement confirms that Transformer is skilled at characterizing deep features of images, and a thorough exploration is worthy.\\
	\noindent\textbf{The effectiveness of DuDoTrans.} When comparing (d) and (e), the boosted 0.18 dB proves SRT effectiveness again. Further, comparing (b) and (e), both dual-domain architectures with corresponding CNNs and Transformer, the improvement demonstrates that Transformer is very suitable in CT reconstruction.
	
	We then investigate the impact of each sub-module on the performance of DuDoTrans:\\
	\noindent\textbf{RIRM depth and width.} Similar to the SRT structure, the RIRM depth represents the number of sub-modules of RIRM, and the RIRM width describes the number of successive Swin-Transformers in each sub-module. Results in Fig~\ref{ablation} (a) and (b) show the corresponding effect of the RIRM width and RIRM depth on the reconstruction performance. When increasing the RIRM depth (with fixed RIRM width 2), the performance is improved quickly when RIRM depth is smaller than 4. Then the PSNR improvement slows down, while the introduced computational cost is increased. Then we fix RIRM depth equal to 3 (blue) and 4 (yellow) and increase RIRM width, and find that the performance is improved fast till RIRM width = 3. After balancing the computation cost and performance improvement, we set RIRM width and depth to 4 and 2, respectively, which is a small model with similar FLOPs to FBPConvNet. \\ 
	\noindent\textbf{The SRT size.} With a similar procedure, we explore the most suitable architecture for the SRT module. As Fig~\ref{ablation} (c) shows, with fixed RIRM depth and width. the performance is not influenced when we enlarge the SRT size (depth $n$ and width $m$ as introduced in Section~\ref{srt}). Specifically, we test five paired models whose RIRM depth and width are set to \{(2, 1), ( 3, 1), (3, 2), (4, 2), (5, 2)\}, respectively. Then we increase ($m$,$n$) from (3, 1) to (4, 2), but the PSNR is even reduced sometimes. Therefore, we set SRT depth and with to (3, 1) as default in later experiments.
	
	\begin{table*}[t!]
		\begin{center}
			\caption{Quantitative results on NIH-AAPM dataset. Our DuDoTrans achieves the best results in all cases. The inferring time is tested when $\alpha_{max}$ is fixed as 96.}
			\vspace{-5pt}
			\label{quantitative-result}
			\resizebox{\linewidth}{!}{
				\begin{tabular}{  l | r |  rr | rr | rr | rr| r }
					\hline\multirow{2}{*}{{NIH-AAPM}}&\multirow{2}{*}{{Param(M)}} & \multicolumn{2}{c|}{{ $\alpha_{max}$ = 24}}  &\multicolumn{2}{c|}{{$\alpha_{max}$ = 72}}   &\multicolumn{2}{c|}{{$\alpha_{max}$ = 96}}  & \multicolumn{2}{c|}{{$\alpha_{max}$ = 144}}  &\multirow{2}{*}{{Time(ms)}} \\ 
					\cline{3-10}
					&& PSNR & SSIM& PSNR &SSIM&PSNR& SSIM &PSNR& SSIM&\\ \hline
					FBP~\cite{natterer2001mathematics} & --  &14.58  & 0.2965  & 17.61 & 0.5085 &18.13   &0.5731     & 18.70  & 0.6668   &  --\\ 
					FBPCovNet~\cite{jin2017deep} & 13.39 &27.10 &0.8158 &30.80 & 0.8671 &31.47 &0.8878 & 32.74   & 0.9084 &155.53 \\  
					DuDoNet~\cite{lin2019dudonet}   &25.80 &26.47 &0.7977&30.94 &0.8816 &31.57  &0.8920   &32.96 &0.9106 &145.65 \\ 
					%					PDNet~\cite{adler2018learned} &  &&&&& &          &          &          &       \\  
					ImgTrans &0.22&\underline{27.46}& \underline{0.8405}&\underline{31.76}&\underline{0.8899}&\underline{32.50}&\underline{0.9010}&\underline{33.50}&\underline{0.9157}&225.56\\
					DuDoTrans &0.44&\textbf{27.55}&\textbf{0.8431} &\textbf{31.91} &\textbf{0.8936} &\textbf{32.68}&\textbf{0.9047}&\textbf{33.70}&\textbf{0.9191}&243.81\\
					\hline
				\end{tabular}
			}
		\end{center}
		\vspace{-10pt}
	\end{table*}
	
	\begin{table*}[t!]
		\begin{center}
			\caption{We test the robustness of DuDoTrans with varied Poisson noise levels. As follows, the intensity is varied to $1{e}^{6}$ (H1), $5{e}^{5}$ (H2), and $1{e}^{5}$, respectively. DuDoTrans keeps the best performance except when the Poisson noise level is enlarged to $1{e}^{5}$, which is too hard to restore clean sinograms.}
			\vspace{-5pt}
			\label{nih-h0}
			\resizebox{\linewidth}{!}{
				\begin{tabular}{  l |  rr | rr | rr | rr| r }
					\hline\multirow{2}{*}{{Noise-H1}}& \multicolumn{2}{c|}{{ $\alpha_{max}$ = 24}}  &\multicolumn{2}{c|}{{$\alpha_{max}$ = 72}}   &\multicolumn{2}{c|}{{$\alpha_{max}$ = 96}}  & \multicolumn{2}{c|}{{$\alpha_{max}$ = 144}}  &\multirow{2}{*}{{Time(ms)}} \\ 
					\cline{2-9}
					& PSNR & SSIM& PSNR &SSIM&PSNR& SSIM &PSNR& SSIM&\\ \hline
					FBP~\cite{natterer2001mathematics} &14.45&0.2815&17.52&0.4898&18.04&0.5541&18.63&0.6483& -- \\ 
					FBPCovNet~\cite{jin2017deep} &27.12& 0.8171&30.74&0.8798&31.44&0.8874&32.65&0.9070&148.58 \\  
					DuDoNet~\cite{lin2019dudonet}   &26.40& 0.7932&30.84 &0.8792 &31.47&0.8900&32.87&0.9090&146.36 \\ 
					ImgTrans &\underline{27.35}&\underline{0.8395}&\underline{31.65}&\underline{0.8882}&\underline{32.42} &\underline{0.8993}&\underline{33.36}&\underline{0.9133}&244.64\\
					DuDoTrans &\textbf{27.45}&\textbf{0.8411}&\textbf{31.80}&\textbf{0.8911}&\textbf{32.55}&\textbf{0.9021}&\textbf{33.48}&\textbf{0.9156}&242.38\\
					\hline
					%					\hline\multirow{2}{*}{{Noise-H2}}& \multicolumn{2}{c|}{{ $\alpha_{max}$ = 24}}  &\multicolumn{2}{c|}{{$\alpha_{max}$ = 72}}   &\multicolumn{2}{c|}{{$\alpha_{max}$ = 96}}  & \multicolumn{2}{c|}{{$\alpha_{max}$ = 144}}  &\multirow{2}{*}{{Time(ms)}} \\ 
					%					\cline{2-9}
					%					& PSNR & SSIM& PSNR &SSIM&PSNR& SSIM &PSNR& SSIM&\\ \hline\hline
					\hline
					\multicolumn{9}{l}{{ Noise-H2}}\\
					\hline
					
					FBP~\cite{natterer2001mathematics} &14.29&0.2652&17.40&0.4688 &17.94&0.5325&18.55&0.6267& --\\ 
					FBPCovNet~\cite{jin2017deep} &27.11 & 0.8168&30.61&0.8764&31.40&0.8865&32.52&0.9047&151.39 \\  
					DuDoNet~\cite{lin2019dudonet}   &26.28&0.7857&30.68&0.8755&31.34&0.8871&32.73&0.9066&152.11 \\ 
					ImgTrans &\underline{27.18}&\underline{0.8361}&\underline{31.49}&\underline{0.8855}&\underline{32.26}&\underline{0.8963}&\underline{33.15}&\underline{0.9096}&244.16\\
					DuDoTrans &\textbf{27.29}&\textbf{0.8377}&\textbf{31.64}&\textbf{0.8881}&\textbf{32.36}&\textbf{0.8986}&\textbf{33.24}&\textbf{0.9113}&261.45\\
					\hline
					%					\hline\multirow{2}{*}{{Noise-H3}}& \multicolumn{2}{c|}{{ $\alpha_{max}$ = 24}}  &\multicolumn{2}{c|}{{$\alpha_{max}$ = 72}}   &\multicolumn{2}{c|}{{$\alpha_{max}$ = 96}}  & \multicolumn{2}{c|}{{$\alpha_{max}$ = 144}}  &\multirow{2}{*}{{Time(ms)}} \\ 
					%					\cline{2-9}
					%					& PSNR & SSIM& PSNR &SSIM&PSNR& SSIM &PSNR& SSIM&\\ \hline\hline
					\hline
					\multicolumn{9}{l}{{ Noise-H3}}\\
					\hline
					
					FBP~\cite{natterer2001mathematics} &13.24&0.1855&16.56&0.3543&17.20&0.4121&17.96&0.5018& --\\ 
					FBPCovNet~\cite{jin2017deep} &\textbf{26.08}&0.7512 & 29.16&0.8294 &30.37&0.8592&31.02&0.8706&152.04 \\  
					DuDoNet~\cite{lin2019dudonet}   &24.77&0.6820 & 29.01&0.8216 &29.86&0.8456&31.25&0.8736& 151.50\\ 
					%				PDNet~\cite{adler2018learned} &  &&&& &          &          &          &       \\  
					ImgTrans &\underline{25.39}&\textbf{0.7933}&\underline{30.20}&\underline{0.8624}&\textbf{31.13} &\underline{0.8707}&\underline{31.52}&\underline{0.8691}&220.78\\
					DuDoTrans &24.77&\underline{0.7844}&\textbf{30.26}&\textbf{0.8632}&\underline{30.86}&\textbf{0.8753}&\textbf{31.58}&\textbf{0.8747}&241.71\\
					\hline
				\end{tabular}
			}
		\end{center}
		\vspace{-10pt}
	\end{table*}
	
	Further, we analyze the convergence, robustness, and effect of the training dataset scale. \\
	\noindent\textbf{Convergence.} In Fig.~\ref{ablation} (d), we plot the convergence curve of FBPConvNet, ImgTrans, and DuDoTrans. Evidently, the introduction of Transformer Structure not only improves the final results, but also stabilizes the training process. Besides, our Dual-Domain design achieves consistently better results, compared with ImgTrans.\\
	\noindent\textbf{Robustness.} In practice, the photon noise in the imaging process influences the reconstructed images, therefore the robustness to such noise is important for application usage.  Here, we simulate such noise with mixed noise (Gaussian \& Poisson noise). Specifically, we train models with default noise level and test them with varied Poisson noise levels (with fixed Gaussian noise), whose intensity correspond to  [1${e}^{\text{5}}$, 5${e}^{\text{5}}$, 1${e}^{\text{6}}$, 5${e}^{\text{6}}$], and show the results in Fig.~\ref{ablation} (e). Evidently, our models achieve better performances except when the intensity is 1${e}^{\text{5}}$, which noise is extremely hard to be suppressed, while DuDoTrans is still better than CNN-based methods, which confirms its robustness.\\	
	\noindent\textbf{Training dataset scale.} Vision Transformers need large-scale data to exhibit performance, and thus limits its development in medical imaging. To investigate it, we train FBPConvNet, ImgTrans, and DuDoTrans with [20\%, 40\%, 60\%, 80\%, 100\%] of our original training dataset, and show the performance in Fig.~\ref{ablation} (f). Obviously, reconstruction performance of DuDoTrans is very stable till the training dataset decreases to 20\%, in which case training data is too less for all models to perform well and DuDoTrans still achieves the best performance.
	\begin{figure*}[!t]
		\begin{minipage}[t]{0.16\textwidth}
			\centering
			\includegraphics[width=\textwidth]{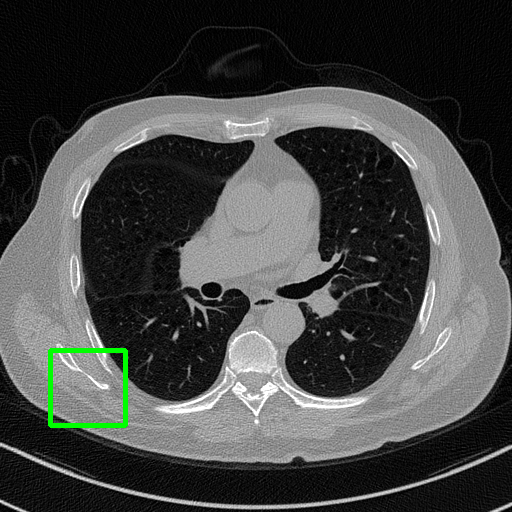}
			\includegraphics[width=\textwidth]{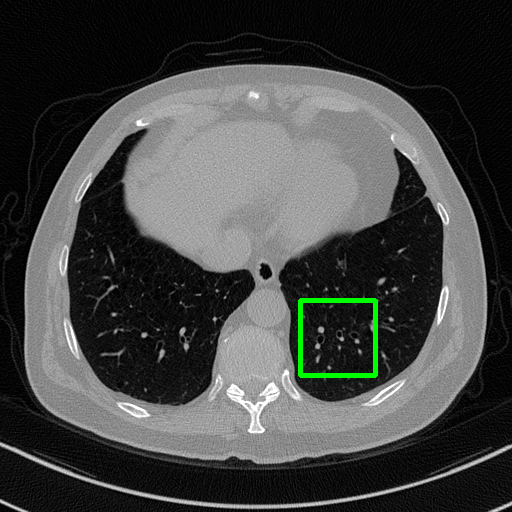}
			\includegraphics[width=\textwidth]{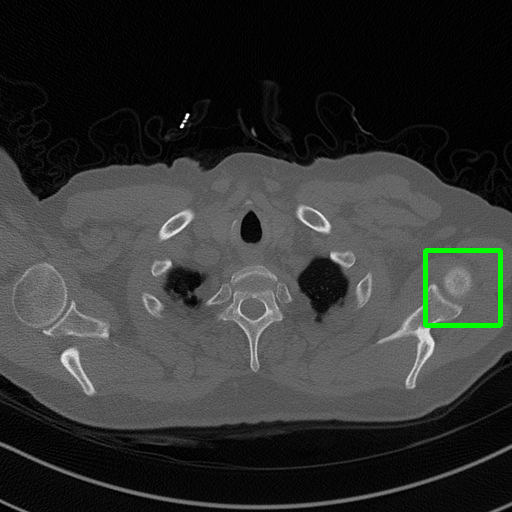}
			Ground Truth
		\end{minipage}
		\begin{minipage}[t]{0.16\textwidth}
			\centering
			\includegraphics[width=\textwidth]{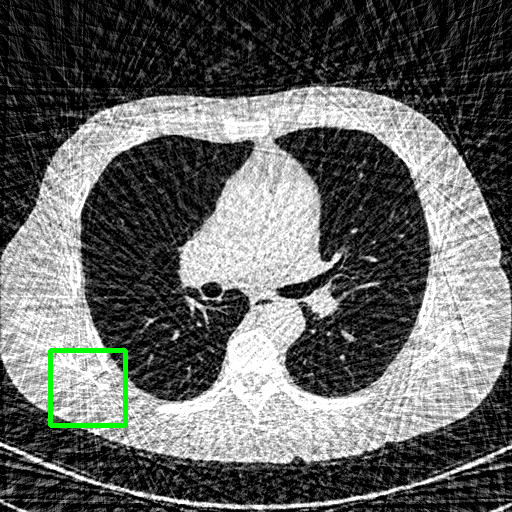}
			\includegraphics[width=\textwidth]{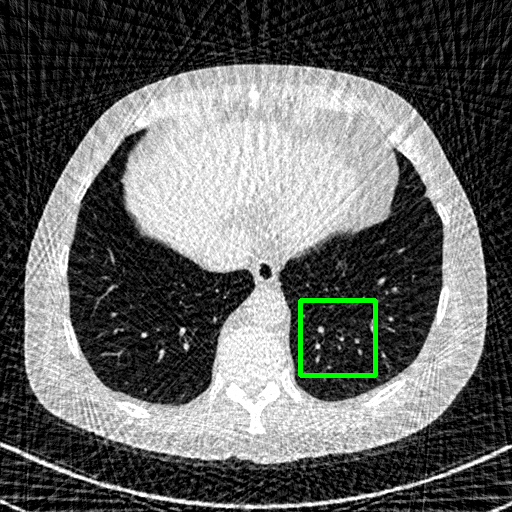}
			\includegraphics[width=\textwidth]{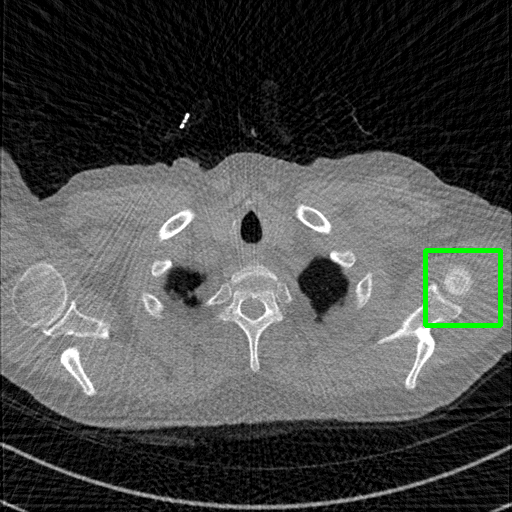}
			FBP
		\end{minipage}
		\begin{minipage}[t]{0.16\textwidth}
			\centering
			\includegraphics[width=\textwidth]{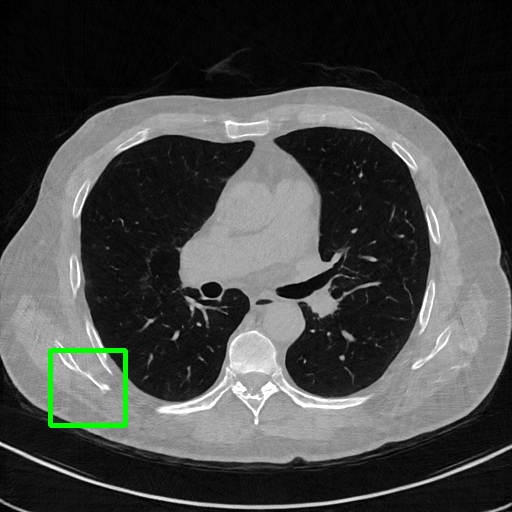}
			\includegraphics[width=\textwidth]{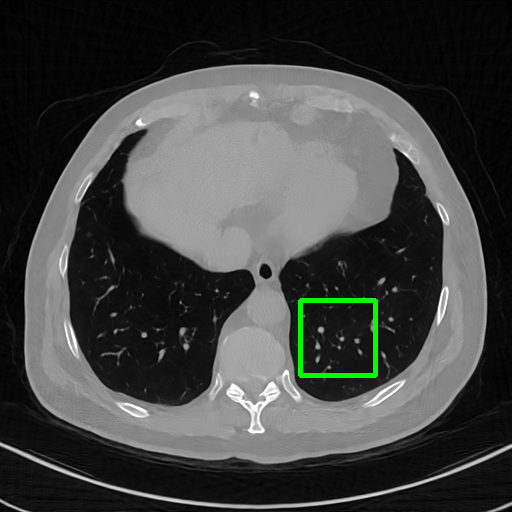}
			\includegraphics[width=\textwidth]{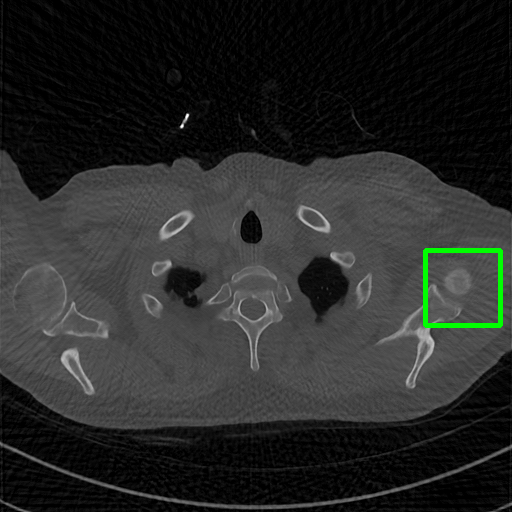}
			FBPConvNet
		\end{minipage}
		\begin{minipage}[t]{0.16\textwidth}
			\centering
			\includegraphics[width=\textwidth]{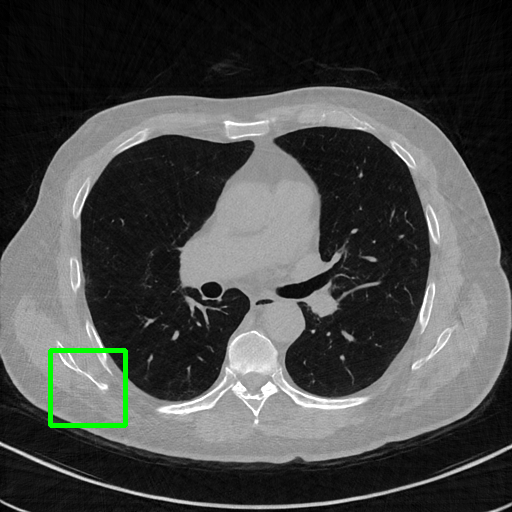}
			\includegraphics[width=\textwidth]{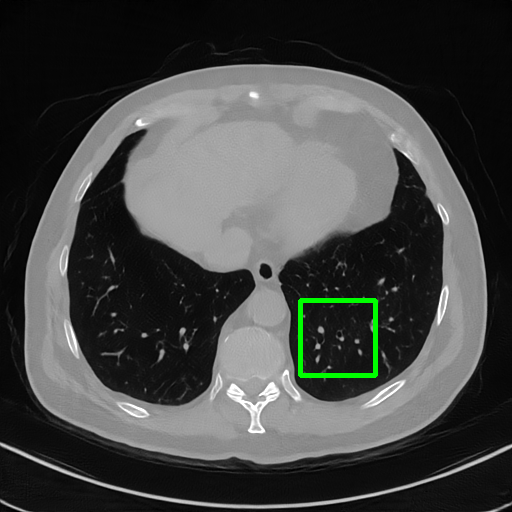}
			\includegraphics[width=\textwidth]{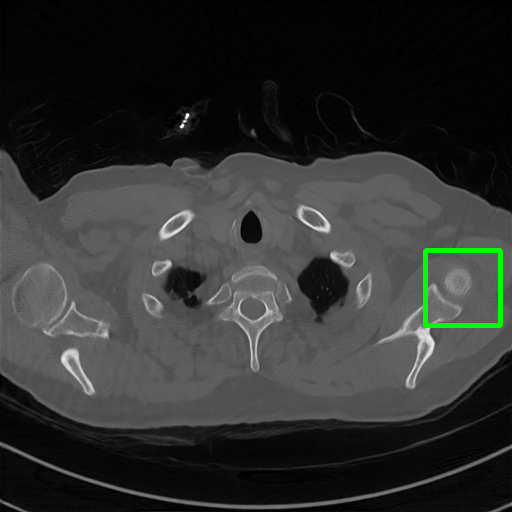}
			DuDoNet
		\end{minipage}
		\begin{minipage}[t]{0.16\textwidth}
			\centering
			\includegraphics[width=\textwidth]{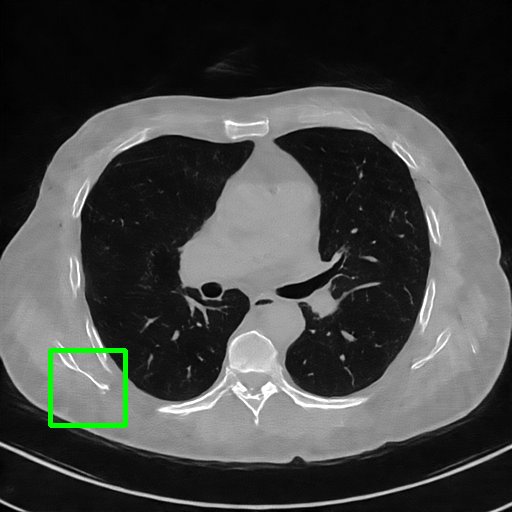}
			\includegraphics[width=\textwidth]{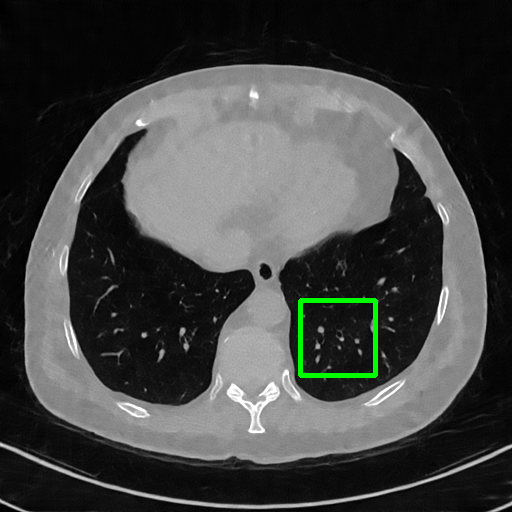}
			\includegraphics[width=\textwidth]{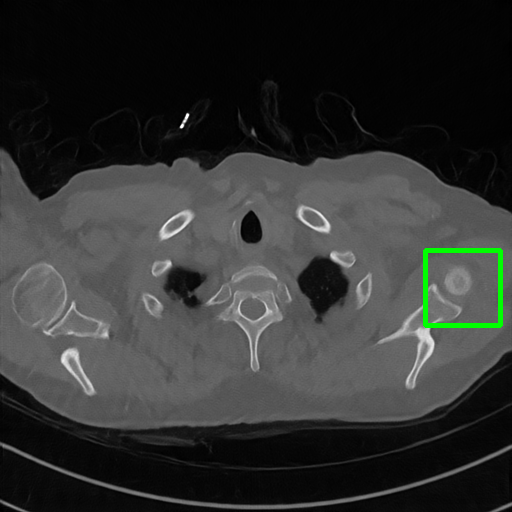}
			ImgTrans
		\end{minipage}
		\begin{minipage}[t]{0.16\textwidth}
			\centering
			\includegraphics[width=\textwidth]{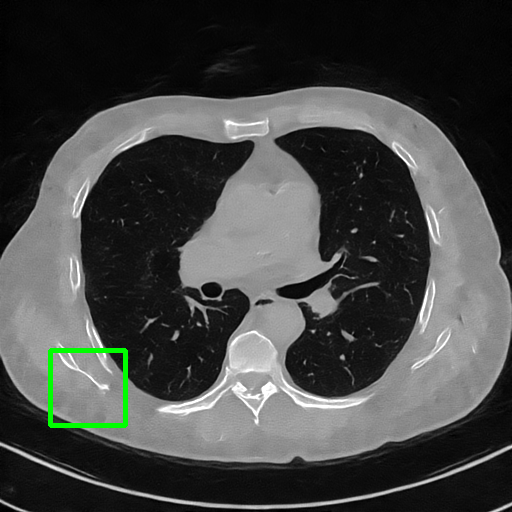}
			\includegraphics[width=\textwidth]{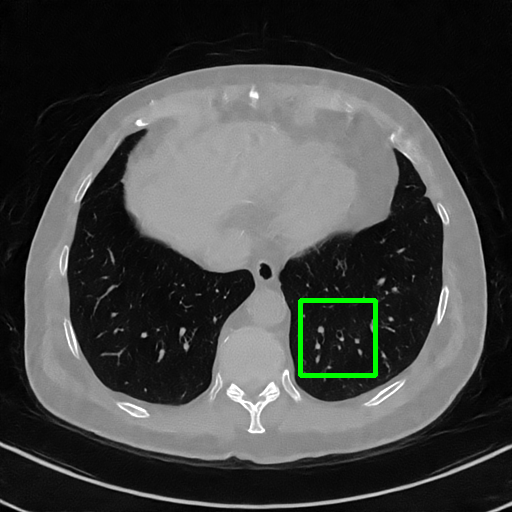}
			\includegraphics[width=\textwidth]{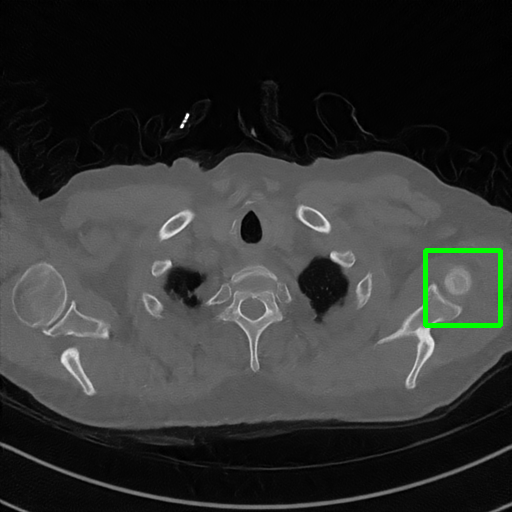}
			DuDoTrans
		\end{minipage}
		\caption{Qualitative comparison on NIH-AAPM dataset, each row from top to bottom corresponds to SV Reconstruction with ${\alpha}_{max}$ = [72, 96 ,144]. The display window is [-1000, 800] HU. We outline the improvements of images with green bounding boxes, and obviously, improvements on visualization are more clear when ${\alpha}_{max}$ increases. }	
		\label{vis-result}
		\vspace{-5pt}
	\end{figure*}
	\begin{table*}[t!]
		\begin{center}
			\caption{Testing performances on COVID-19 dataset. Given these unobserved data, our DuDoTrans still performs the best in all cases.}
			\label{covid}
			\resizebox{\linewidth}{!}{
				\begin{tabular}{ l |  rr | rr | rr | rr| r }
					\hline\multirow{2}{*}{{COVID-19}}& \multicolumn{2}{c|}{{ $\alpha_{max}$ = 24}}  &\multicolumn{2}{c|}{{$\alpha_{max}$ = 72}}   &\multicolumn{2}{c|}{{$\alpha_{max}$ = 96}}  & \multicolumn{2}{c|}{{$\alpha_{max}$ = 144}}  &\multirow{2}{*}{{Time(ms)}} \\ 
					\cline{2-9}
					& PSNR & SSIM& PSNR &SSIM&PSNR& SSIM &PSNR& SSIM&\\ \hline
					FBP~\cite{natterer2001mathematics} & 14.82 & 0.3757 &18.16 &0.5635 &18.81 &0.6248  &19.36 & 0.7070 & --\\  
					FBPCovNet~\cite{jin2017deep} & 26.43 & 0.8015 & 32.84 & 0.9407 &  33.72 & 0.9409     & 34.62 & 0.9651 &149.48 \\  
					DuDoNet~\cite{lin2019dudonet} &26.97&0.8558&33.10 &0.9429 & 32.57 &0.9380   &36.13 &0.9722 & 153.25\\ 
					%				PDNet~\cite{adler2018learned} &&&&&& &          &          &          &            \\ 
					ImgTrans      &\underline{27.24}&\underline{0.8797}&\underline{35.58}&\underline{0.9580}&\underline{37.31}&\underline{0.9699}&\underline{39.90}&\textbf{0.9801}&222.14\\ 
					DuDoTrans &\textbf{27.74}&\textbf{0.8897}&\textbf{35.62}&\textbf{0.9596}&\textbf{37.83}&\textbf{0.9727}&\textbf{40.20}&\underline{0.9794}&244.46\\
					\hline
				\end{tabular}
			}
		\end{center}
		\vspace{-20pt}
	\end{table*}
	
	\subsection{Sparse-View CT Reconstruction Analysis}
	We next conduct thorough experiments to test the performance of DuDoTrans on various sparse-view scenarios. Specifically, we first train models when ${\alpha}_{max}$ is 24, 72, 96, 144, respectively. The results are shown in Table~\ref{nih-h0}, and DuDoTrans have achieved consistently better results. Besides, we observe that ImgTrans and DuDoTrans are more stable in training, and the learned parameters are both extremely small, compared with CNN-based models. Furthermore, the improvement of DuDoTrans over ImgTrans becomes larger when the ${\alpha}_{max}$ increases, which confirms the usefulness of restored sinograms in reconstruction. \\
	\noindent\textbf{Qualitative comparison.} We also visualize the reconstructed images of these methods in Fig.~\ref{vis-result} with ${\alpha}_{max}$ = [72, 96 ,144] (See more visualizations in Appendix). In all three rows, our DuDoTrans shows better detail recovery, and sparse-view artifacts are suppressed. Further, when decreasing ${\alpha}_{max}$, where raw sinograms are too messy to be restored and low-quality images from FBP are too hard to capture global features, Transformer-based models exhibit reduced performance. The phenomena suggests that we should design suitable structures with the Transformer and CNNs, facing with different cases. \\
	
	\begin{figure*}[t!]
		\begin{minipage}[t]{0.16\textwidth}
			\centering
			\includegraphics[width=\textwidth]{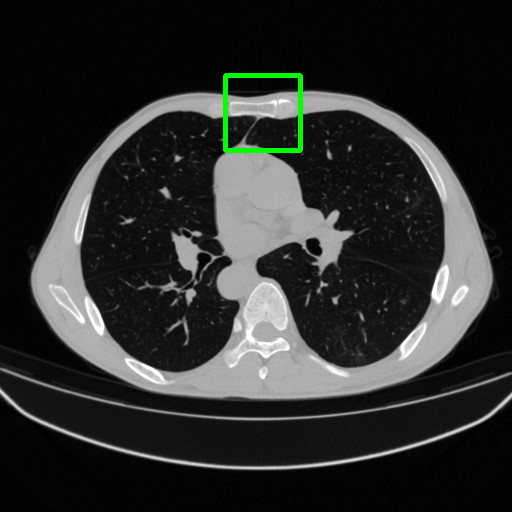}
			Ground Truth
		\end{minipage}
		\begin{minipage}[t]{0.16\textwidth}
			\centering
			\includegraphics[width=\textwidth]{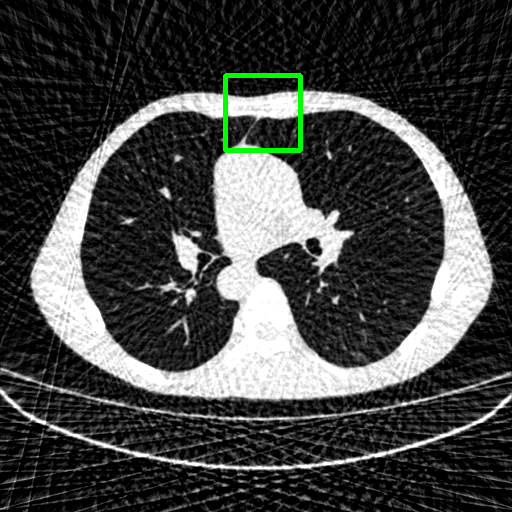}
			FBP
		\end{minipage}
		\begin{minipage}[t]{0.16\textwidth}
			\centering
			\includegraphics[width=\textwidth]{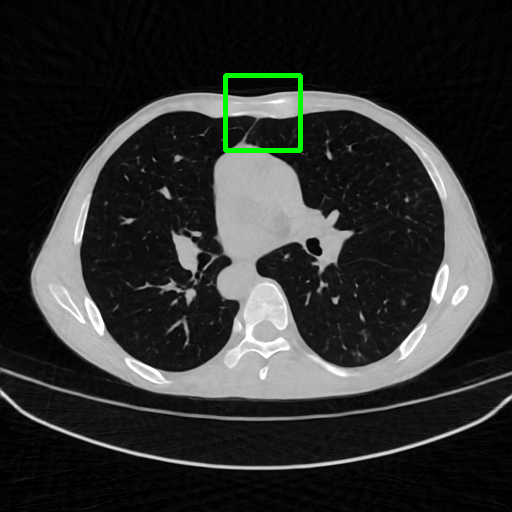}
			FBPConvNet
		\end{minipage}
		\begin{minipage}[t]{0.16\textwidth}
			\centering
			\includegraphics[width=\textwidth]{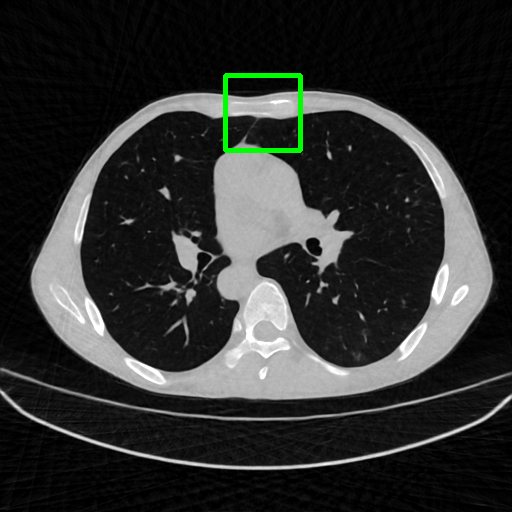}
			DuDoNet
		\end{minipage}
		\begin{minipage}[t]{0.16\textwidth}
			\centering
			\includegraphics[width=\textwidth]{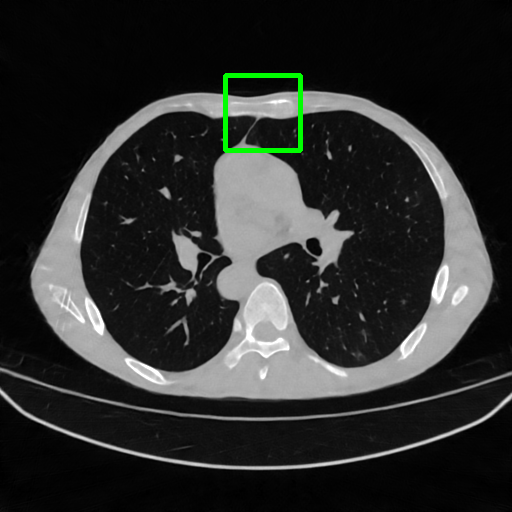}
			ImgTrans
		\end{minipage}
		\begin{minipage}[t]{0.16\textwidth}
			\centering
			\includegraphics[width=\textwidth]{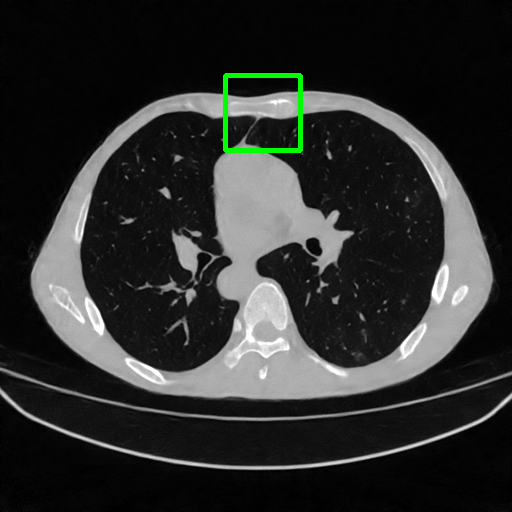}
			DuDoTrans
		\end{minipage}
		\vspace{-5pt}
		\caption{Qualitative comparison on COVID-19 dataset when  ${\alpha}_{max}$ = 96. The display window is [-1000, 800] HU. We outline the improvements of images with green bounding boxes, which shows that DuDoTrans performs better that others.}
		\label{vis-result-covid}
		\vspace{-15pt}
	\end{figure*}
	
	\begin{figure}[t]
		\begin{center}
			\includegraphics[height=6.0cm]{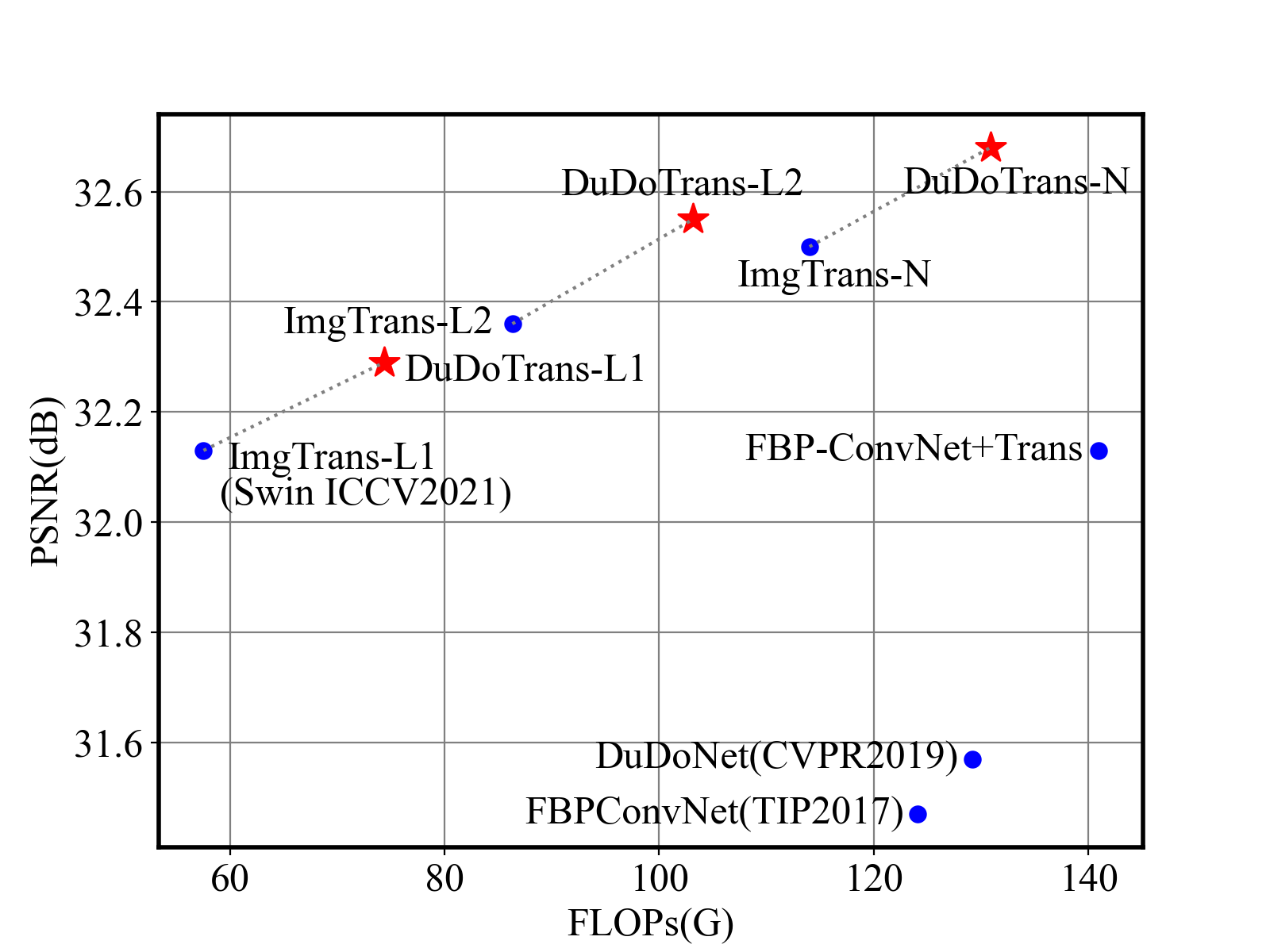}
			\caption{FLOPs versus performances of these methods.  Similar to Fig~\ref{performance_params}, L1, L2 are two light versions and N represents the normal version. ImgTrans-L1 and DuDoTrans-L1 have achieved 0.5-0.7 dB improvements with less than 80G FLOPs, while CNN-based methods needs over 120G FLOPs. Further, DuDoTrans-N has enlarged the improvement to 1.2 dB with similar FLOPs.}
			\label{performance_flops}
		\end{center}
		\vspace{-25pt}
	\end{figure}
	
	\vspace{-17pt}
	\noindent\textbf{Robustness with noise in SVCT.} With the decrease of view number ${\alpha}_{max}$, input sinograms would be messier, which makes SVCT more difficult. Therefore, we test the robustness of all trained models with aforementioned Poisson noise levels when ${\alpha}_{max}$ = [24, 72, 96, 144], and report performances in Table.~\ref{nih-h0}. The included notation Noise-H1, Noise-H2, and Noise-H3 correspond to Poisson intensity  [1${e}^{\text{6}}$, 5${e}^{\text{5}}$, 1${e}^{\text{5}}$]. Compared with CNN-based methods, ImgTrans and DuDoTrans show better robustness. Across involved cases, DuDoTrans shows the best performance. Nevertheless, when Poisson intensity is 1${e}^{\text{5}}$, DuDoTrans fails to exceed ImgTrans and FBPConvNet, which is caused by the extremely messy sinograms. In this case, restoring sinograms is too difficult and DuDoNet also fails.\\    
	\noindent\textbf{Generalizablity on COVID-19 dataset.} Finally, we use slices of another patient in the COVID-19 dataset to test the generalizability of trained models, and quantitative performances are compared in Table~\ref{covid}. ImgTrans and DuDoTrans have achieved a larger improvement about 4-5 dB over CNN-based methods, which shows that the long-range dependency modeling ability helps capture the intrinsic global property of general CT images. Further, our DuDoTrans exceeds ImgTrans about 0.4 dB in all cases, even larger than the original NIH-AAPM dataset. The improvement ensures that DuDoTrans generalizes well to out-of-distribution CT images. Besides, we also show the visualization images in Fig.~\ref{vis-result-covid} when ${\alpha}_{max}$ = 96. Coinciding with the quantitative comparison, our DuDoTrans show better reconstruction on both global patterns and local details.	
	
	\subsection{Computation comparison}
	As a practical problem, reconstruction speed is necessary when deployed in modern CT machines. Therefore, we compare the parameters and FLOPs versus performances in Fig~\ref{performance_params} and Fig.~\ref{performance_flops}, respectively. We find that Transformer-based methods have achieved better performances with fewer parameters, and our DuDoTrans exceeds ImgTrans with only a few additional parameters. As known, the patch-based operations and the attention mechanism are computationally expensive, which limits their application usage. Therefore, we further compare the FLOPs of these methods. As shown, light versions (DuDoTrans-L1, DuDoTrans-L2) have achieved 0.8-1 dB improvement with fewer FLOPs, and DuDoTrans-N with default size has enlarged the improvement to 1.2 dB. Besides, we report the inferring time in each Table~\ref{quantitative-result}~\ref{nih-h0}~\ref{covid}, and computation time is very similar to CNN-based methods, whose additional consumption is because of the patch-based operations.
	
	\begin{table}[t!]
		\begin{center}
			\caption{Test performances of various models w/ v.s. w/o SRT module. Obviously, SRT improves performances of CNN-based, Transformer-based, and Deep unrolloing methods. Note that we replace RIRM with PDNet in our framework, named PDNet+SRT.}
			\label{srt-improve}
			\begin{tabular}{ l | l |rr} 
				\hline 
				&	Method                  &PSNR & SSIM  \\
				\hline
				
				&	FBPConvNet~\cite{jin2017deep}         & 31.47 & 0.8878  \\
				w/o SRT	&	PDNet~\cite{adler2018learned}                   &31.62  & 0.8894 \\
				&	ImgTrans  &  32.50&  0.9010\\
				\hline
				
				&	\textbf{FBPConvNet+SRT}& \textbf{32.13}   & \textbf{ 0.8989} \\
				w/ SRT	&	\textbf{PDNet+SRT}        & \textbf{32.38}  &\textbf{ 0.9045}\\
				&	\textbf{DuDoTrans}    &\textbf{32.68} &\textbf{0.9047}  \\
				\hline
			\end{tabular}
		\end{center}
		\vspace{-20pt}
	\end{table}
	
	\subsection{Discussion}
	%As known, the advantage of Transformer is long-range dependency modeling, while the local feature extraction ability of CNNs is powerful. Although the Transformer-based structure show sota performance in vision tasks, fewer works analyze its advantage in medical imaging. In our experiments, Transformer-based structures, ImgTrans and DuDoTrans, exhibit better stability, robustness, and higher performance when ${\alpha}_{max}$ is [72, 96, 144]. While the performance of these two methods reduces in extremely sparse-view scenarios. We analyze this is because the low-quality images from FBP provide few global patterns, which are disrupted by the induced artifacts. This case suggests that we should design suitable architectures for different problems. Besides, the proposed SRT is shown effective when combined with both FBPConvNet and ImgTrans, which confirms the matching between Transformer and sinogram restoration. We will also explore the combination of SRT with iterative and deep-unrolling methods in our future work to extend our proposed framework.
	
	As in Table~\ref{effect-srt}, we have shown the effectiveness of SRT with FBPConvNet~\cite{jin2017deep}  and ImgTrans~\cite{liu2021swin}, which are two post-processing methods. Recently deep-unrolling methods have attracted much attention in reconstruction. To concretely verify the SRT module's effectiveness, we further combine it with PDNet~\cite{adler2018learned}, which is a deep-unrolling method. Results of involved three paired models (w/ v.s. w/o SRT) are shown in Table~\ref{srt-improve} with default experimental settings when ${\alpha}_{max}$ = 96 (See other cases when ${\alpha}_{max}$ = [24, 72, 144] in Appendix). All these three reconstruction methods have been improved by the use of the SRT module. Furthermore, our DuDoTrans still performs the best without any unrolling design. Thus, our SRT is flexible and powerful probably in any existing reconstruction framework.
	
	\section{Conclusion}
	We propose a transformer-based SRT module with long-range dependency modeling capability to exploit the global characteristics of sinograms, and verify it in CNN-based, Transformer-based, and Deep-unrolling reconstruction framework. Further, via combining SRT and similarly-designed RIRM, we yield DuDoTrans for SVCT reconstruction. Experimental results on the NIH-AAPM dataset and COVID-19 dataset show that DuDoTrans achieves state-of-the-art reconstruction. To further benefit DuDoTrans with the accordingly designing advantage of deep-unrolling methods, we will explore ``DuDoTrans + unrolling" in the future.
	% better than CNN-based methods, and the dual-domain design improves ImgTrans by about 0.2-0.4 dB.	
	
	\vspace{0.3cm}
	\noindent \textbf{Acknowledge.}
	This work was supported by the National Natural Science
	Foundation of China under Grant No. 12001180
	and 12101061.
%	\clearpage
	%%%%%%%%% REFERENCES
	{\small
		\bibliographystyle{ieee_fullname}
		\bibliography{egbib}
	}
	
\end{document}